\documentclass[12pt,apj]{emulateapj}
\usepackage{graphicx}
\usepackage{epsfig}
\usepackage{epstopdf}
\usepackage{natbib}
\begin{document}

\title{What causes the high apparent speeds in chromospheric and transition region spicules on the Sun?}
\shortauthors{De Pontieu et al.}
\shorttitle{High Apparent Speeds in Solar Spicules}

\author{Bart De Pontieu \altaffilmark{1,2} \& Juan Mart\'inez-Sykora \altaffilmark{1,3} \& Georgios Chintzoglou \altaffilmark{1,4} }
\email{bdp@lmsal.com}
\affil{\altaffilmark{1} Lockheed Martin Solar and Astrophysics Laboratory, Palo Alto, CA 94304, USA}
\affil{\altaffilmark{2} Institute of Theoretical Astrophysics, University of Oslo, P.O. Box 1029 Blindern, N-0315 Oslo, Norway}
\affil{\altaffilmark{3} Bay Area Environmental Research Institute, Petaluma, CA 94952, USA}
\affil{\altaffilmark{4} University Corporation for Atmospheric Research, Boulder, CO 80307-3000, USA} 
\newcommand{\eg}{{\it e.g.,}} 
\newcommand{\myemail}{bdp@lmsal.com}
\newcommand{\komment}[1]{\texttt{#1}}
\newcommand{\ul}{\underline}
\newcommand{\pref}{\protect\ref}
\newcommand{\soho}{{\em SOHO{}}}
\newcommand{\sdo}{{\em SDO{}}}
\newcommand{\stereo}{{\em STEREO{}}}
\newcommand{\iris}{{\em IRIS{}}}
\newcommand{\hinode}{{\em Hinode{}}}
\newcommand{\B}{$\bullet$ }
\newcommand{\si}{\ion{Si}{4}~1402\AA }
\newcommand{\sir}{\ion{Si}{4}}
\newcommand{\ct}{\ion{C}{2}~1335\AA }
\newcommand{\ctr}{\ion{C}{2}}
\newcommand{\fen}{\ion{Fe}{9}~171\AA }
\newcommand{\fern}{\ion{Fe}{9}}
\newcommand{\fet}{\ion{Fe}{12}~193\AA }
\newcommand{\fert}{\ion{Fe}{12}}
\newcommand{\fef}{\ion{Fe}{14}~211\AA }
\newcommand{\ferf}{\ion{Fe}{14}}
\newcommand{\ly}{Ly$\alpha$}
\newcommand{\longacknowledgment}{IRIS is a NASA small explorer mission developed and operated by LMSAL with mission operations executed at NASA Ames Research center and major contributions to downlink communications funded by ESA and the Norwegian Space Centre. We gratefully acknowledge support by NASA grants
NNX11AN98G, NNX16AG90G, and NASA contracts NNM07AA01C (Hinode), and NNG09FA40C (IRIS). The simulations have been run on clusters from the Notur project, 
and the Pleiades cluster through the computing project s1061 from the High 
End Computing (HEC) division of NASA. 
Snapshots from this numerical simulation (en096014\_gol) are publicly available as part of the \iris\ modeling archive. Details can be found at http://iris.lmsal.com/modeling.html . To analyze the data we have used IDL.
}

\begin{abstract}
Spicules are the most ubuiquitous type of jets in the solar atmosphere.
The advent of high-resolution imaging and spectroscopy from the Interface Region Imaging Spectrograph (IRIS) and 
ground-based observatories has revealed the presence of very high apparent motions 
of order 100-300 km~s$^{-1}$ in spicules, as measured in the plane of the sky. 
However, line-of-sight measurements of such high speeds have been difficult to obtain, with values deduced from Doppler shifts in spectral lines typically of order 30-70 km~s$^{-1}$.
In this work we resolve this long-standing discrepancy using recent 2.5D radiative MHD simulations. This simulation has revealed a novel driving mechanism for spicules 
in which ambipolar diffusion resulting from ion-neutral interactions plays a key role. In our simulation we often see that
the upward propagation of magnetic waves and electrical currents from the low chromosphere into already existing spicules 
can lead to rapid heating when the currents are rapidly dissipated by ambipolar diffusion. The combination of rapid heating and the propagation of these currents at Alfv\'enic speeds in excess of 100 km~s$^{-1}$
leads to the very rapid apparent motions, and often wholesale appearance, of spicules at chromospheric and transition region temperatures. In our simulation, the observed fast apparent motions in such jets 
are actually a signature of a heating front, and much higher than the mass flows, which are of order 30-70 km~s$^{-1}$. Our results can explain the behavior of transition region ``network jets'' and the very high apparent speeds reported for some chromospheric spicules.


\end{abstract}


\section{Introduction}
Jets are very common in the solar chromosphere. They are most often detected as rapidly evolving linear features in which plasma appears to be accelerated and sometimes heated, with the bulk of the plasma flow often penetrating into the overlying and much hotter corona. There is a wide variety of chromospheric jets, including surges, penumbral microjets, anemone jets, macrospicules and spicules \citep{Raouafi:2016fe}. To fully understand the impact of these different jets on the mass and energy balance of the low solar atmosphere, it is key to better understand the mechanisms that drive these jets. In this paper we focus on the most common of solar jets, spicules, as they have the largest potential to impact the overlying corona or solar wind \citep{Beckers:1968qe,Athay:1982fk,McIntosh:2011fk,De-Pontieu:2011lr}. During the past decade, significant progress has been made in better understanding spicules and how they impact the solar atmosphere, but key questions remain. 

At least two different types of spicules have been reported. The so-called type I spicules, identified as dynamic fibrils or mottles on the disk, show both apparent motions and mass flows of maximally 10-40~km~s$^{-1}$ and appear to be driven by magneto-acoustic shocks that form when convective motions, p-modes or magnetic disturbances propagate upward into the chromosphere \citep{Hansteen+DePontieu2006,De-Pontieu:2007cr,luc2007, Martinez2009}. The transition region response to type I spicules appears to be limited to brightenings at the top of the spicule \citep{Skogsrud2016}. The bulk of spicules observed at the solar limb are however type II spicules. These are much faster, showing apparent motions with maximum speeds of 30-150~km~s$^{-1}$ \citep{de-Pontieu:2007kl,Pereira2012}. Such speeds dominate the spicules seen in the chromospheric \ion{Ca}{2} H passband, which recent results indicate are the initial upward phase of longer-lived spicules that show up-and-downward motion when viewed over a wide range of chromospheric and TR passbands \citep{Skogsrud2015}. In contrast to type I spicules, they often appear to be associated with heating to at least transition region temperatures \citep{De-Pontieu2014b,Rouppe2015}. Their short-lived ($\sim 20-50$s) transition region counterparts often appear in \sir\ and \ctr\ slit-jaw images from the Interface Region Imaging Spectrograph \citep[IRIS,][]{De-Pontieu:2014yu} as linear features that emanate from magnetic network with even higher apparent speeds of 80-300 km~s$^{-1}$ \citep{Tian:2014fp, Narang2016}. The transition region counterparts appear to be shorter and slower in quiet Sun than in coronal holes, presumably because of differing magnetic field configurations \citep{Narang2016}. Such high apparent speeds have also been found in other upper chromospheric lines such as \ly\ \citep{Kubo:2016lq}. There is however a large discrepancy between these high apparent speeds and the actual mass flows in spicules as deduced from Doppler shift measurements. For example, the on-disk chromospheric counterparts of type II spicules are rapid blueshifted events (RBEs) which show typical Dopplershifts of 20-50 km~s$^{-1}$ \citep{Sekse2012,Sekse2013}. Similarly, Doppler shift measurements in transition region lines of spicules show velocities up to only 50-70 km~s$^{-1}$ \citep{Rouppe2015}. 

The viewing geometry could in principle explain some of the differences between the apparent motions, often measured at the limb, and Doppler shifts, typically measured on the disk. However, the fact is that Doppler shifts of 100-300 km~s$^{-1}$ are extremely uncommon on the disk and would be expected if the plane-of-the-sky motions were real mass motions. More importantly, the \iris\ observations of the apparent and real mass motions of transition region counterparts of spicules \citep{Tian:2014fp,Rouppe2015} are both on the disk, i.e., with similar viewing angles. These differences thus appear to be real and not easy to explain, thus providing strict constraints for any theoretical model of spicules. A multitude of theoretical models have been developed in the past, but until recently no single model has been able to simultaneously explain, in detail, the ubiquity, dynamic and thermal evolution and visibility and appearance in various chromospheric and transition region observables \citep[for a review, see][]{Tsiropoula2012,Sterling2000}. Here we exploit a recent numerical model based on 2.5D radiative MHD simulations that proposes that type II spicules occur when magnetic tension, created through the interaction between strong flux concentrations and weaker, granular scale, magnetic flux concentrations, can diffuse into the chromosphere through ambipolar diffusion. In the chromosphere, the sudden release of tension drives strong flows and ion-neutral interactions lead to heating of the spicular plasma \citep{Martinez2017}. We compare the model predictions with observations from IRIS.


\begin{figure*}[tbh]
	\begin{center}
           \includegraphics[width=0.95\hsize]{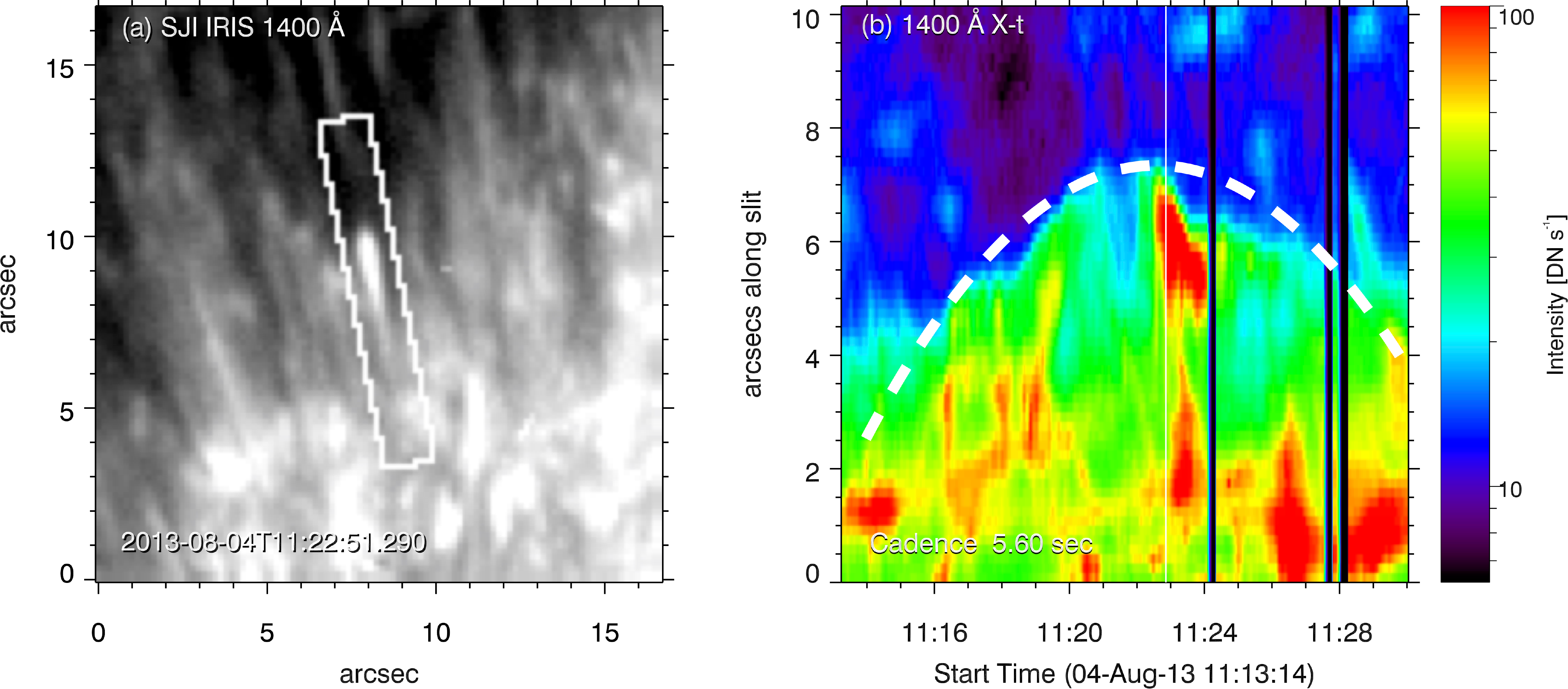}		
          \includegraphics[width=0.95\hsize]{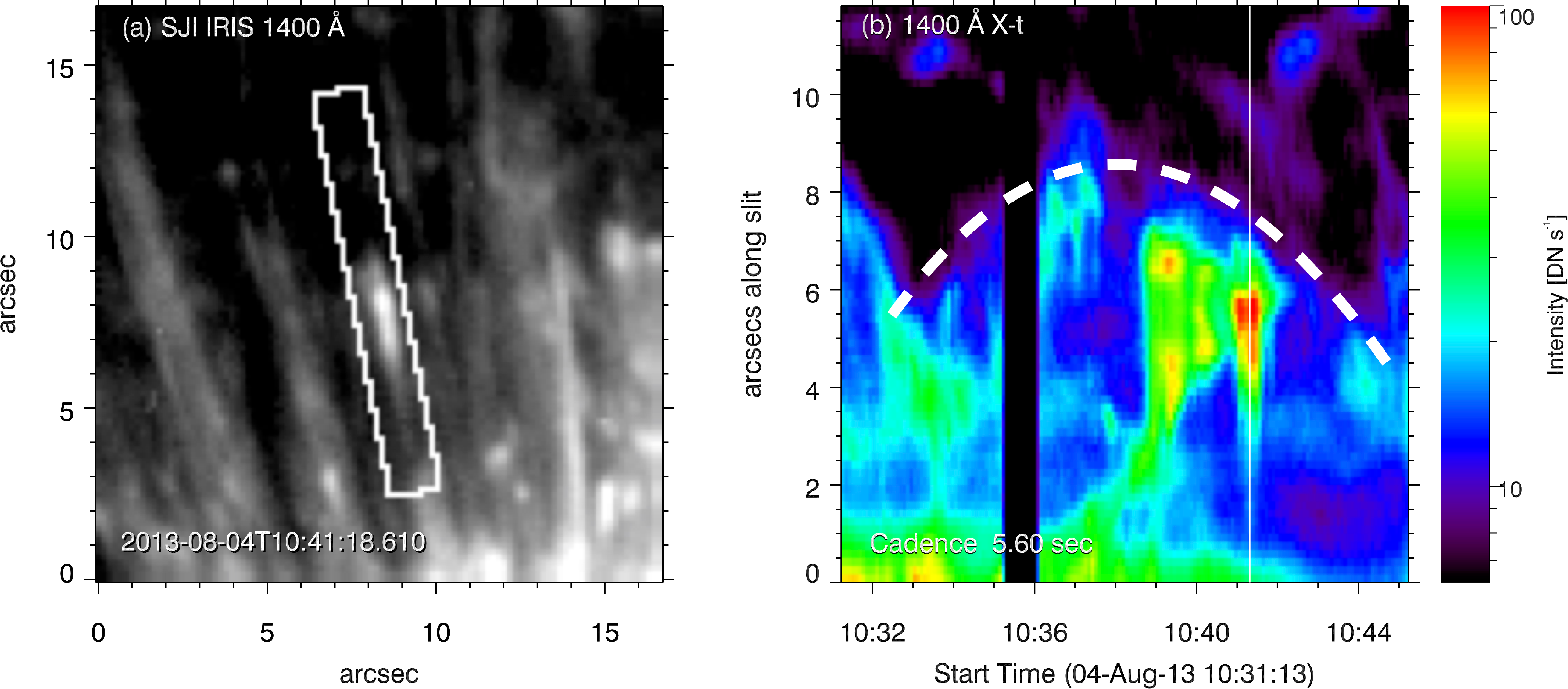}
		\caption{The space-time evolution of \sir\ from observations with the \iris\ 1400 slit-jaw channel reveals linear, jet-like, intensity brightenings (red, almost vertical features in the right panels) at t=10:41:00 UTC (bottom) and 11:22:20 UTC (top) that propagate at high speeds along an already existing spicule (which is outlined as a parabolic shape in the right panels). The left panels show maps of the SJI 1400 and the rectangles highlight the spicules. For the space-time plot (right panels), we follow the intensity (in IRIS Data Numbers per second) along the major axis of the rectangles in the left plot. The black vertical lines in the space-time plots are caused by datagaps in the time series. This figure is accompanied by two on-line animations.} 
		\label{fig:obs}
	\end{center}
\end{figure*}

\section{Observations}~\label{sec:obs}

We use IRIS 1400\AA\ slit-jaw images (SJI) of an on-disk plage region (NOAA AR 11809) centered 
at $[x,y]= [383,124]$\arcsec, with a Field-Of-View of $119\arcsec \times 120$\arcsec. The observations
were taken on 04-Aug-2013 10:38-11:41 UTC. IRIS was in sit-and-stare mode (IRIS OBS-ID 4043007648) 
taking only the 1400\AA\ SJI passband images with cadence of 5.6~s and exposure time of 4~s. 
We use level-2 IRIS data that was corrected for flat-field, dark current, geometry and co-alignment
 \citep{De-Pontieu:2014yu}.

\begin{figure}[tbh]
\begin{center}
\includegraphics[width=0.8\hsize]{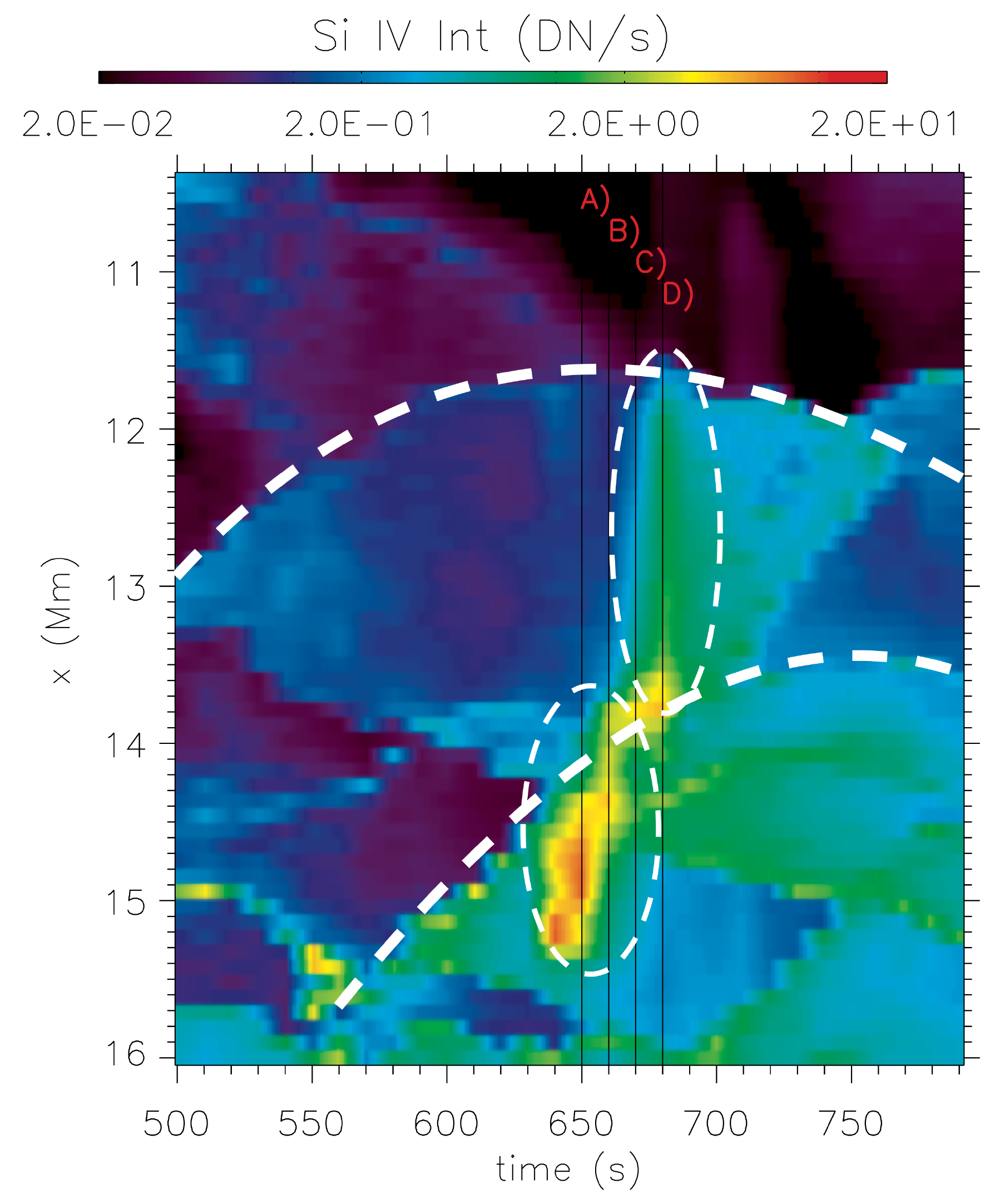}
\caption{Space-time evolution of a top-view of \sir\ (in IRIS Data Numbers per second assuming the signal would cover an \iris~pixel of 0.166\arcsec x 0.166\arcsec) from a 2.5D radiative MHD simulation reveals two jet-like features (inside dashed white ovals) that propagate at high apparent speeds at t=630~s ($x=14-15.5$ Mm) and at t=680~s ($x=11.5-13.5$ Mm), with similar speeds as seen in the observations (Fig.~\ref{fig:obs}). These jets form on already existing spicules (whose space-time path is outlined with dashed white parabolas). The vertical 
solid lines are the instances shown in Figures~\ref{fig:tgevo} and~\ref{fig:current}. This plot is based on a top-view of the simulation shown in Fig.~\ref{fig:tgevo}, i.e., the x-coordinates are the same in both figures.} 
\label{fig:sievo}
\end{center}
\end{figure}

\section{Simulations}~\label{sec:sim}

We compare our observations with a numerical simulation that uses Bifrost, a radiative MHD code \citep{Gudiksen:2011qy}.
This code solves the MHD equations including radiative transfer from 
the photosphere to corona \citep{Hayek:2010ac,carlsson:2012uq} and 
thermal conduction along the magnetic field. We also include ion-neutral interaction effects by using the Generalized Ohm's law, i.e., by adding the ambipolar diffusion and Hall term to the induction equation 
\citep{Martinez-Sykora:2012uq,Martinez2017b}. 

\begin{figure*}[tbh]
\begin{center}
\includegraphics[width=0.96\hsize]{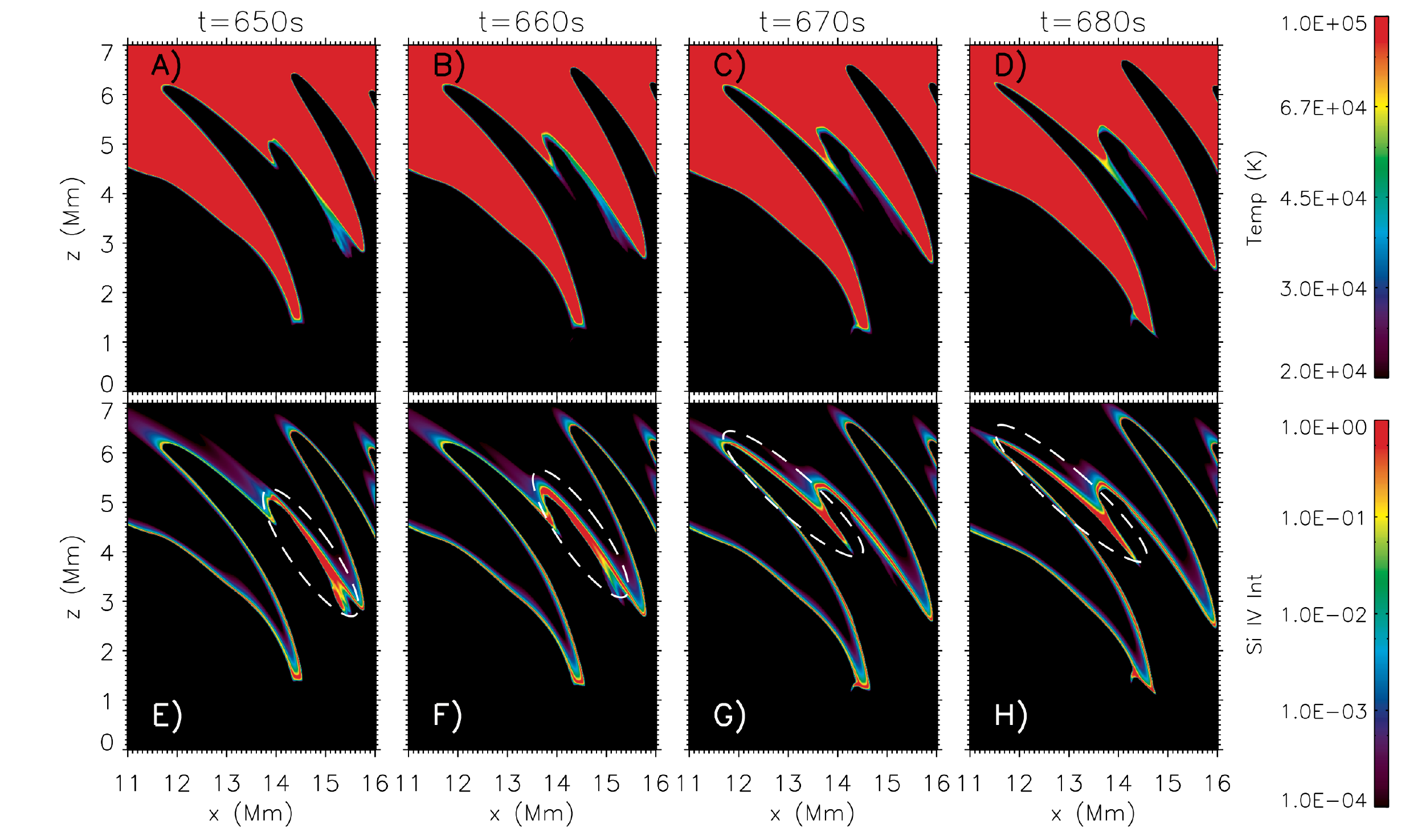}
\caption{Maps of temperature (top row), and \si\ emissivity (bottom row) from our numerical simulation
reveal very fast thermal evolution and accompanying increase in \si\ intensity along already existing spicules. 
The time interval between the snapshots is only 10~s. The dashed ovals highlight the increased \ion{Si}{4} emission that is associated with the propagation and dissipation of electrical currents and subsequent heating of threads in the spicule. These threads cause the almost vertical bright features (dashed white ovals) in Fig.~\ref{fig:sievo}. This figure is accompanied by two on-line animations, one showing the temporal evolution of the figure, and one blinking the temperature and \si\ emissivity plots.} 
\label{fig:tgevo}
\end{center}
\end{figure*}

The simulation analyzed here is the 2.5D radiative MHD model described in detail 
in \citet{Martinez2017b, Martinez2017}.  The numerical domain ranges from the upper layers 
of the convection zone ($3$~Mm below the photosphere) to the self-consistently maintained hot corona ($40$~Mm above the 
photosphere) and $90$~Mm along the horizontal axis. We use this particular model because it is the first Bifrost simulation in wich 
type II spicules occur ubiquitously. \citet{Martinez2017} show that ambipolar diffusion plays a dominant role in the formation of spicules. To compare our simulation with observations, we calculate the synthetic \sir\ intensity assuming equilibrium ionization and the optically thin approximation, similar to the methods used by \citet{Hansteen:2010uq}. 


\section{Results}~\label{sec:res}

Measurements of apparent motions of spicules in the plane-of-the-sky reveal a wide range of velocities, with the highest velocities recorded in \iris\ slit-jaw images. These are sensitive to emission from lower transition region lines like \ct\ (SJI 1330 \AA) and \si\ (SJI 1400 \AA). Previous results indicate that velocities of 80-250 km~s$^{-1}$ \citep{Tian:2014fp} are common. A statistical study shows that apparent velocities of up to 350 km~s$^{-1}$ are sometimes seen, with clear differences between coronal hole ($190\pm60$ km~s$^{-1}$) and quiet Sun ($110\pm40$ km~s$^{-1}$) spicules \citep{Narang2016}. Such fast motions in spicules also occur in active regions: Figure~\ref{fig:obs} shows two examples of spicules (left panels) that emanate from strong magnetic flux concentrations in plage as observed in the \iris\ SJI 1400~\AA\ channel. The space-time evolution (right panels) of these two different spicules show similar behavior. We see very rapid (i.e., almost vertical in the space-time plot) linear brightenings (red in the right panels) that shoot outward from the spicule footpoints. These fast apparent jets appear to occur along an existing spicule and cover the full length of the spicule ($\sim4$\arcsec, i.e., 3~Mm) within a few tens of seconds. The previously existing spicule is seen as a very faint (note the logarithmic intensity scale in the right panels) parabolic path with a lifetime of $\sim10$ (top) and 8 (bottom) minutes. Most spicules do not show such clear parabolic profiles in space-time plots derived from IRIS slitjaws. In those cases, the only observed features are the rapid apparent jets. This is especially the case in quiet Sun and coronal hole spicule observations. The fast apparent jets are sometimes repetitive, with linear brightenings recurring along the same spicules (bottom panel of Figure~\ref{fig:obs}). 

\begin{figure*}[tbh]
\begin{center}
\includegraphics[width=0.96\hsize]{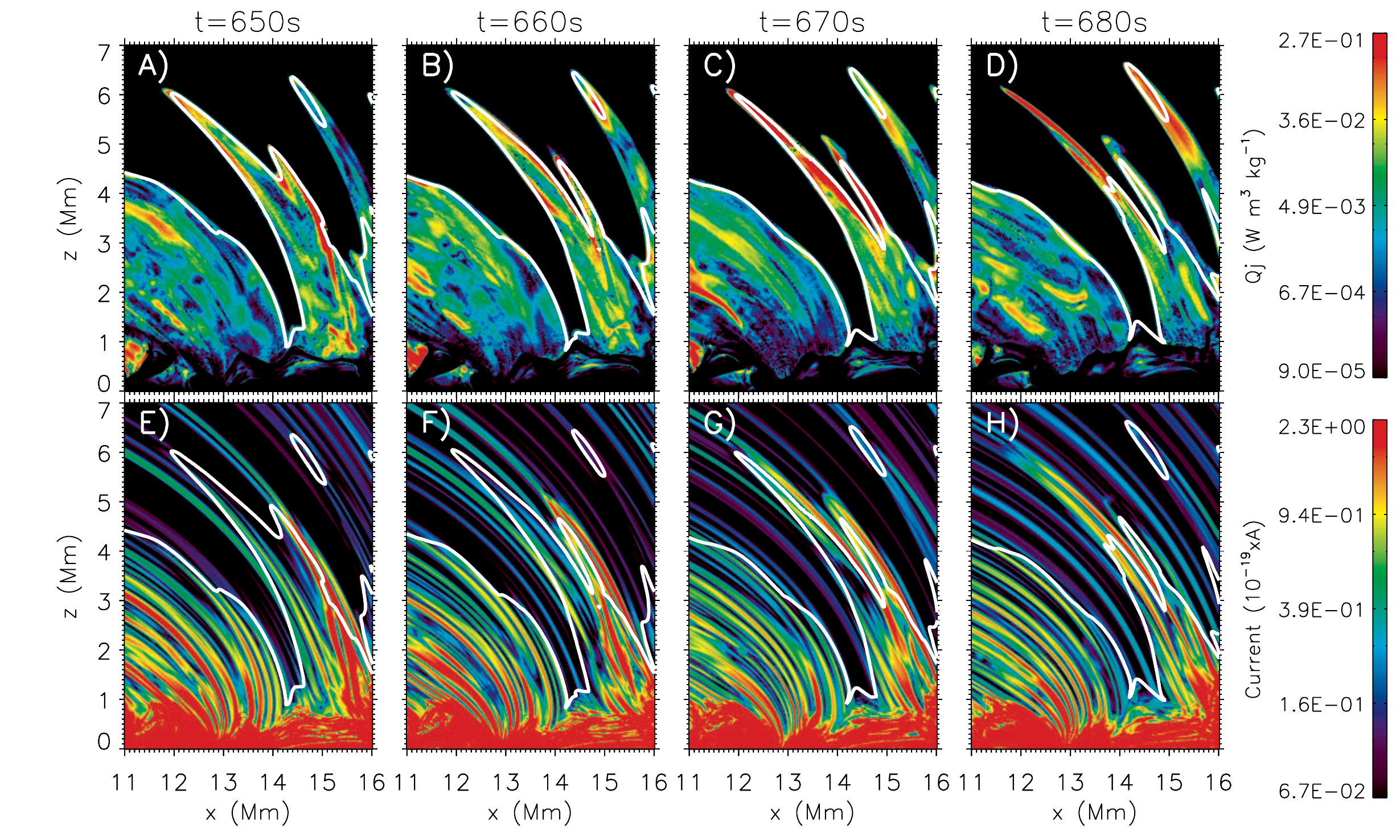}
\caption{Maps of Joule heating due to the ambipolar diffusion (top row), and current
density (bottom row) from our numerical simulation reveal currents being propagated 
at Alfv\'enic speed and dissipated along already existing spicules. 
The interval between the snapshots is only 10~s. The temperature at 
$10^4$~K is shown with white contours.} 
\label{fig:current}
\end{center}
\end{figure*}

Similar behavior is seen in the synthetic transition region observables calculated from our numerical simulation.
Upper chromospheric and transition region observables reveal the presence of linear features with very fast apparent motions associated with the simulated spicules.
The example in the space-time plot of synthetic \si\ intensity in Figure~\ref{fig:sievo} reveals two very fast apparent jets that appear almost consecutively, one next to the other. These apparent jets occur on an already fully evolved spicule which is ``filled'' along its whole length with a strong \si\ brightening within 10 seconds. The first event is 
at $x=[13.9,15.4]$~Mm, and $t\sim650$~s and the second event is at $x=[11.2,13.6]$~Mm, and $t\sim680$~s. When seen from above (top view along the negative z-axis), these linear features would show speeds of roughly 150-200~km~s$^{-1}$. When viewed from a line-of-sight that is perpendicular to the simulated spicules, the apparent motions would be even faster (200-300 km~s$^{-1}$). Our simulations show that the apparent jets (observed in TR lines) occur around the time of the maximum extent of the chromospheric spicule (visible as parabolic paths overdrawn in Fig.~\ref{fig:sievo}), i.e., when the chromospheric mass flows are already much reduced (of order $<20$~km~s$^{-1}$) compared to the very high chromospheric mass flows (up to 70 km~s$^{-1}$) during the initial acceleration of the jet. At the time of the occurrence of the TR counterparts, their apparent speeds are thus much higher than the chromospheric mass flows.


The cause for these apparent \si\ jets is elucidated by Figure~\ref{fig:tgevo} which shows the temporal evolution of the 
temperature and the synthetic \si\ intensity. In the center of the domain ($x=12-14.5$~Mm), we can see two fully developed spicules that undergo very rapid thermal changes. The  spicule located on the left hand side ($x\sim12-14$~Mm) shows an increase of \sir\ intensity from its footpoint towards the top, all along the right side during the time range $t=660-680$~s. This sudden increase in transition region spectral line intensity is caused by a thermal perturbation that propagates upward along the spicule and travels roughly 3~Mm in about 10 to 20~s, i.e., with a propagation speed of 150-300~km~s$^{-1}$ along the spicule (Figure~\ref{fig:tgevo}). This heating front is localized in a very narrow region within the spicule and leads to a significant increase in the \si\ intensity. The propagation speed of the heating front is much faster than the actual mass flows at this time. The spicule on the right hand side in Fig.~\ref{fig:tgevo} similarly shows a rapidly propagating heating front that causes the earlier apparent jet that is visible at $x=14-16$ Mm in Fig.~\ref{fig:sievo}.

\begin{figure*}[tbh]
	\begin{center}
		\includegraphics[width=0.45\hsize]{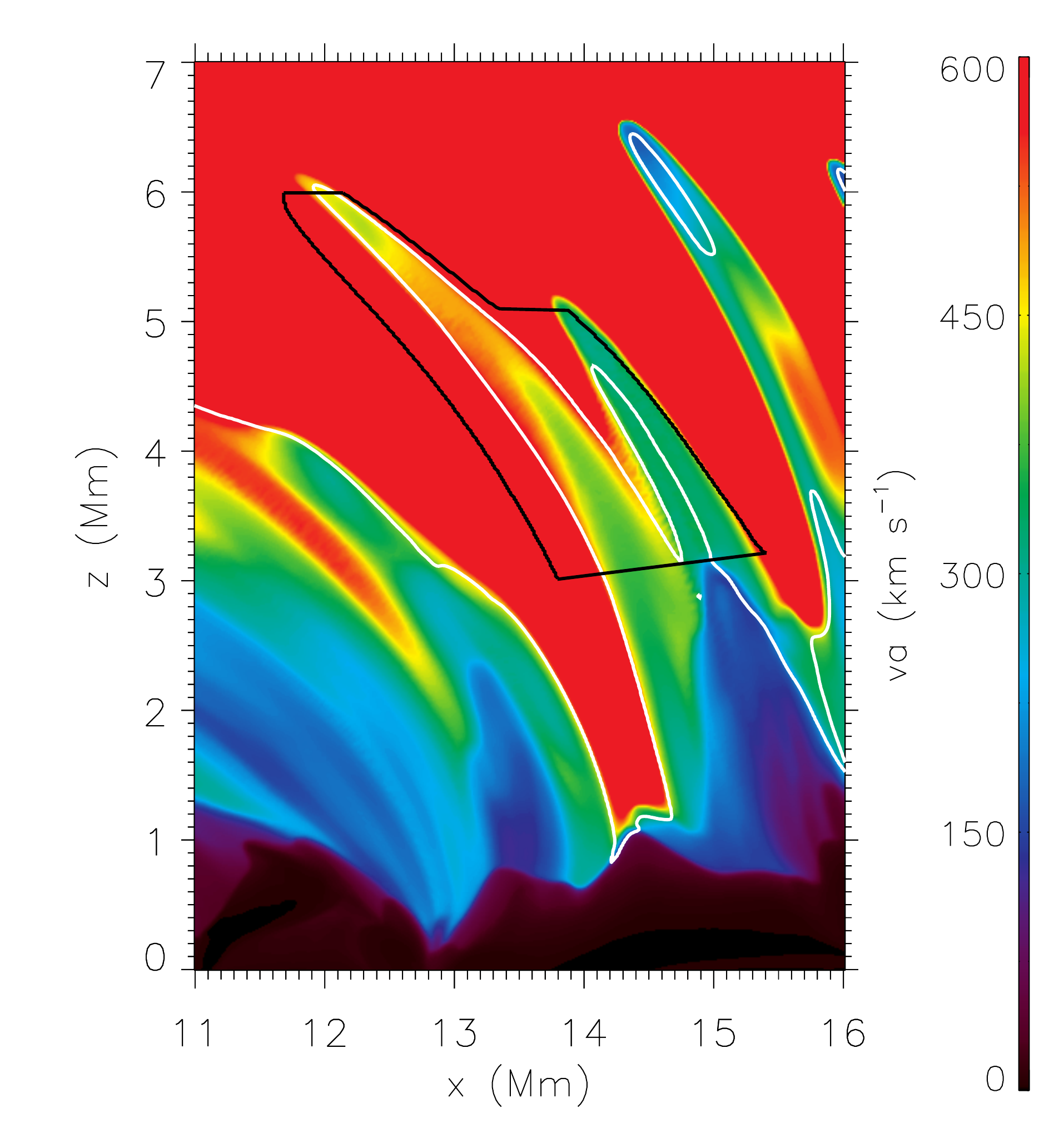}
		\includegraphics[width=0.65\hsize]{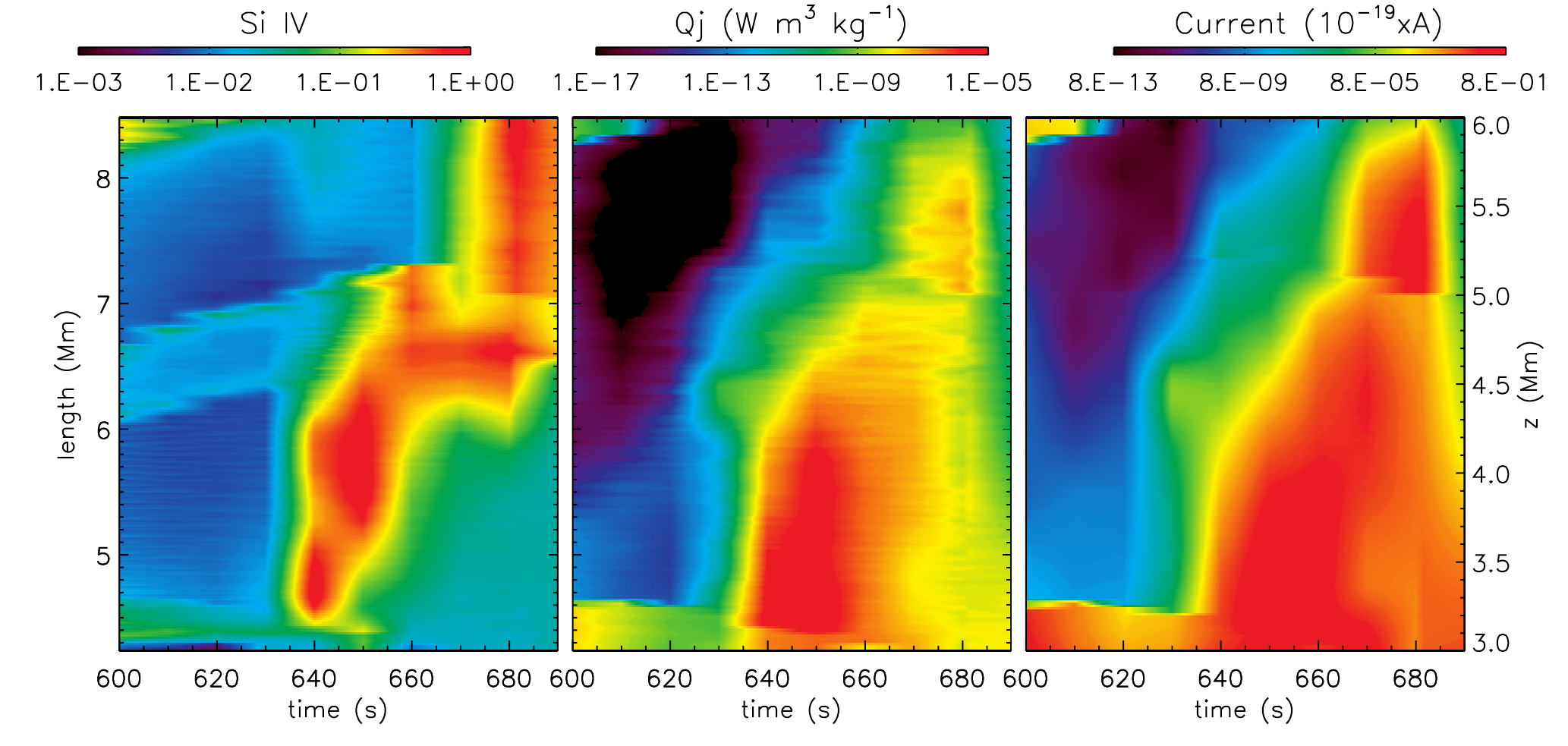}
		\caption{The heating fronts in the simulated spicules propagate with the Alfv\'en speed. The top shows a map of the Alfv\'en speed at $t=660$~s (i.e., the same simulation 
		snapshot as in Panels B and F in Figures~\ref{fig:tgevo} and~\ref{fig:current}) which has values of between 150 and 450 km~s$^{-1}$ in the spicule. The white contour shows locations with temperatures of $10^4$~K. The black contour shows the region over which we sum (at each height) to calculate the space-time plots in the bottom row. The bottom row shows, as a function of time, the \si\ emissivity, Joule heating rate and electrical current along the length of the spicule as measured from the photosphere (left vertical axis, for equivalent height, see right vertical axis). The heating front in the right spicule dominates the signal around $t=640$~s, while the heating front in the left spicule (at greater heights) occurs around $t=680$~s. We see a clear correlation between the rapid \si\ increase in both ``jets'' and Alfv\'enic propagation of electrical current and associated dissipation. 
} 
		\label{fig:va}
	\end{center}
\end{figure*}

In our simulation, the appearance of these linear features with high apparent propagation speeds is caused by the following mechanism.
After the simulated type II spicules are already fully formed, the low density spicular environment causes the collision frequency between ions and neutrals to drop, resulting in strong ambipolar diffusion along the spicules. Meanwhile, in the photosphere below the spicule, magnetic energy is built up due to the convective motions. Large currents in the upper photosphere are seen in red in the bottom row of Figure~\ref{fig:current}. Eventually, these currents escape into the chromosphere. For instance, see the current on right hand side of the left spicule between $t=[660,680]$~s. These electric currents propagate along the magnetic field at Alfv\'enic speeds, driven by tension and/or transverse waves. The map of Alfv\'en speeds shown in the top panel of Figure~\ref{fig:va} reveals values between $150-450$~km~s$^{-1}$ within the spicule. Since the ambipolar diffusion is so high along the spicule, the magnetic energy is dissipated on very short times-scales (a few seconds). The heating from ambipolar diffusion is shown in the top row of Figure~\ref{fig:current}. This strong heating increases plasma temperatures from chromospheric ($\sim8\times 10^3$~K) to upper-chromospheric or transition region temperatures ($10^4- 10^5$~K) on short time scales, i.e., between $t=660$ and 680~s for the left spicule. Similar to the intensity and temperature maps shown in Figure~\ref{fig:tgevo}, the heating and current are very collimated and localized in a very narrow region along the spicule ($\sim100~km$). The cause for the high apparent speeds is illustrated further in the bottom row of Fig.~\ref{fig:va} which shows a space-time plot along the length of (and summed across) the two spicules. The first apparent jet occurs in the right spicule (t=630s) and is associated with a strong current and associated heating front that causes a large increase in \si\ intensity. The second apparent jet follows a similar scenario and occurs at t=670s in the left spicule at somewhat greater heights. These time-scales, speeds, and synthetic observables all are in agreement with the observations \citep{Tian:2014fp,Narang2016}. We note that this is not the only possible cause for fast apparent jets at transition region temperatures. As mentioned, the initial violent acceleration of the spicule often involves mass motions of order 70 km~s$^{-1}$: it is thus possible that the low-end of the reported velocity range is associated with real mass motions. However, our results suggest that the bulk of the very high apparent speeds are likely caused by heating fronts that propagate at Alfv\'enic speeds.




\section{Discussion \& conclusions}~\label{sec:dis}

Our results suggest that fast apparent motions seen in linear features that appear to be the transition region counterparts of spicules \citep{Tian:2014fp} are often not caused by real mass motions. Instead these motions in the plane-of-the-sky are of order 100-300 km~s$^{-1}$ and may be caused by the rapid propagation, at Alfv\'enic speeds, of heating fronts. A detailed comparison with our 2.5D radiative MHD simulation indicates that these heating fronts are an integral part of the spicule formation process and caused by the rapid dissipation of electric currents that propagate from the photosphere through the spicule into the corona. The dissipation mechanism is driven by the collisions between ions and neutrals: in the low density spicular environment, the ion-neutral collision frequency is significantly decreased leading to slippage between the ionized and neutral particles. This slippage causes dissipation of currents and subsequent heating of plasma.

Our simulation can explain the well-known mismatch between the speeds measured from imaging timeseries (100-300 km~s$^{-1}$) and spectroscopic measurements of Doppler shifts ($<100$ km~s$^{-1}$). It also suggests that the linear shape of such features, often taken as a signature for the presence of a jet, may be misleading. In our case the linear feature forms because plasma is heated along an already formed jet, but the rapid formation of the linear features is not necessarily associated with the formation of the jet. It means that the observed motions are not caused by mass motions, which impacts the estimated mass and kinetic energy flux that these jets may provide into the solar wind \citep{Tian:2014fp}. Our results not only provide an explanation for the rapid ``network jets'' observed with \iris\ \citep{Tian:2014fp}, but could also provide an explanation for the rapid upper chromospheric signals that have been recently observed with the CLASP rocket instrument \citep{Kubo:2016lq}.

The excellent match between the simulation and observations provides more support for the spicule formation mechanism proposed by \citet{Martinez2017}. This model proposes that spicules form when ambipolar diffusion in the low chromosphere allows magnetic fields, resulting from the interaction between strong network or plage fields and the ubiquitous, weak, granular-scale magnetic fields, to emerge into the upper chromosphere. The subsequent violent release of magnetic tension drives strong mass flows and heating when the associated electric currents are dissipated.  The heating produced by the dissipation of these currents is significant and could potentially play a substantial role in the energy balance of the chromosphere. Further studies will be required to determine whether spicules (or ``straws'') indeed play a substantial role in heating the chromosphere as tentatively suggested by \citet{Cauzzi2009}.

Our results show that there are indeed strong mass motions caused by ejection of spicular material along flux-tube like features. The fast apparent motions in some observables are caused by vigorous heating from the dissipation through ambipolar diffusion of upward propagating electrical currents. Our results thus indicate that chromospheric spicules or their transition region counterparts do not appear to be caused by random superposition or warping of current sheets along the line-of-sight \cite[as suggested previously by][]{Judge2011}.

Despite the complex and large variety of physical processes included in this simulation, it suffers from certain limitations which one must address in further studies. This model is limited to only 2.5 spatial dimensions and assumes that Hydrogen and Helium ionization are in statistical equilibrium, whereas in the chromosphere both are far from statistical equilibrium  \citep{Leenaarts:2007sf,Golding:2014fk}. In addition, the generalized Ohm's law is based on limiting assumptions which may break down in the dynamic spicular environment: a full multi-fluid approach will provide further insight into the role of ion-neutral interactions in spicules. Finally, while the numerical resolution is high enough to properly treat the ambipolar resistivity, the lack of 3D treatment and resolution means that we cannot exclude the possibility that the actual dissipation mechanism driving the heating fronts is intimately connected to dissipation of Alfv\'en waves, e.g., through the Kelvin Helmholtz instability \citep{Antolin2015b}.




\longacknowledgment{}


\bibliographystyle{apj}

\begin{thebibliography}{33}
\expandafter\ifx\csname natexlab\endcsname\relax\def\natexlab#1{#1}\fi

\bibitem[{{Antolin} {et~al.}(2015){Antolin}, {Okamoto}, {De Pontieu},
  {Uitenbroek}, {Van Doorsselaere}, \& {Yokoyama}}]{Antolin2015b}
{Antolin}, P., {Okamoto}, T.~J., {De Pontieu}, B., {Uitenbroek}, H., {Van
  Doorsselaere}, T., \& {Yokoyama}, T. 2015, \apj, 809, 72

\bibitem[{{Athay} \& {Holzer}(1982)}]{Athay:1982fk}
{Athay}, R.~G. \& {Holzer}, T.~E. 1982, \apj, 255, 743

\bibitem[{{Beckers}(1968)}]{Beckers:1968qe}
{Beckers}, J.~M. 1968, \solphys, 3, 367

\bibitem[{{Carlsson} \& {Leenaarts}(2012)}]{carlsson:2012uq}
{Carlsson}, M. \& {Leenaarts}, J. 2012, \aap, 539, A39

\bibitem[{{Cauzzi} {et~al.}(2009){Cauzzi}, {Reardon}, {Rutten}, {Tritschler},
  \& {Uitenbroek}}]{Cauzzi2009}
{Cauzzi}, G., {Reardon}, K., {Rutten}, R.~J., {Tritschler}, A., \&
  {Uitenbroek}, H. 2009, \aap, 503, 577

\bibitem[{{De Pontieu} {et~al.}(2007{\natexlab{a}}){De Pontieu}, {Hansteen},
  {Rouppe van der Voort}, {van Noort}, \& {Carlsson}}]{De-Pontieu:2007cr}
{De Pontieu}, B., {Hansteen}, V.~H., {Rouppe van der Voort}, L., {van Noort},
  M., \& {Carlsson}, M. 2007{\natexlab{a}}, \apj, 655, 624

\bibitem[{{De Pontieu} {et~al.}(2007{\natexlab{b}}){De Pontieu}, {McIntosh},
  {Hansteen}, {Carlsson}, {Schrijver}, {Tarbell}, {Title}, {Shine}, {Suematsu},
  {Tsuneta}, {Katsukawa}, {Ichimoto}, {Shimizu}, \&
  {Nagata}}]{de-Pontieu:2007kl}
{De Pontieu}, B., {McIntosh}, S., {Hansteen}, V.~H., {Carlsson}, M.,
  {Schrijver}, C.~J., {Tarbell}, T.~D., {Title}, A.~M., {Shine}, R.~A.,
  {Suematsu}, Y., {Tsuneta}, S., {Katsukawa}, Y., {Ichimoto}, K., {Shimizu},
  T., \& {Nagata}, S. 2007{\natexlab{b}}, \pasj, 59, 655

\bibitem[{{De Pontieu} {et~al.}(2011){De Pontieu}, {McIntosh}, {Carlsson},
  {Hansteen}, {Tarbell}, {Boerner}, {Martinez-Sykora}, {Schrijver}, \&
  {Title}}]{De-Pontieu:2011lr}
{De Pontieu}, B., {McIntosh}, S.~W., {Carlsson}, M., {Hansteen}, V.~H.,
  {Tarbell}, T.~D., {Boerner}, P., {Martinez-Sykora}, J., {Schrijver}, C.~J.,
  \& {Title}, A.~M. 2011, Science, 331, 55

\bibitem[{{De Pontieu} {et~al.}(2014{\natexlab{a}}){De Pontieu}, {Rouppe van
  der Voort}, {McIntosh}, {Pereira}, {Carlsson}, {Hansteen}, {Skogsrud},
  {Lemen}, {Title}, {Boerner}, {Hurlburt}, {Tarbell}, {Wuelser}, {De Luca},
  {Golub}, {McKillop}, {Reeves}, {Saar}, {Testa}, {Tian}, {Kankelborg},
  {Jaeggli}, {Kleint}, \& {Martinez-Sykora}}]{De-Pontieu2014b}
{De Pontieu}, B., {Rouppe van der Voort}, L., {McIntosh}, S.~W., {Pereira},
  T.~M.~D., {Carlsson}, M., {Hansteen}, V., {Skogsrud}, H., {Lemen}, J.,
  {Title}, A., {Boerner}, P., {Hurlburt}, N., {Tarbell}, T.~D., {Wuelser},
  J.~P., {De Luca}, E.~E., {Golub}, L., {McKillop}, S., {Reeves}, K., {Saar},
  S., {Testa}, P., {Tian}, H., {Kankelborg}, C., {Jaeggli}, S., {Kleint}, L.,
  \& {Martinez-Sykora}, J. 2014{\natexlab{a}}, Science, 346, 1255732

\bibitem[{{De Pontieu} {et~al.}(2014{\natexlab{b}}){De Pontieu}, {Title},
  {Lemen}, {Kushner}, {Akin}, {Allard}, {Berger}, {Boerner}, {Cheung}, {Chou},
  {Drake}, {Duncan}, {Freeland}, {Heyman}, {Hoffman}, {Hurlburt}, {Lindgren},
  {Mathur}, {Rehse}, {Sabolish}, {Seguin}, {Schrijver}, {Tarbell},
  {W{\"u}lser}, {Wolfson}, {Yanari}, {Mudge}, {Nguyen-Phuc}, {Timmons}, {van
  Bezooijen}, {Weingrod}, {Brookner}, {Butcher}, {Dougherty}, {Eder},
  {Knagenhjelm}, {Larsen}, {Mansir}, {Phan}, {Boyle}, {Cheimets}, {DeLuca},
  {Golub}, {Gates}, {Hertz}, {McKillop}, {Park}, {Perry}, {Podgorski},
  {Reeves}, {Saar}, {Testa}, {Tian}, {Weber}, {Dunn}, {Eccles}, {Jaeggli},
  {Kankelborg}, {Mashburn}, {Pust}, {Springer}, {Carvalho}, {Kleint}, {Marmie},
  {Mazmanian}, {Pereira}, {Sawyer}, {Strong}, {Worden}, {Carlsson}, {Hansteen},
  {Leenaarts}, {Wiesmann}, {Aloise}, {Chu}, {Bush}, {Scherrer}, {Brekke},
  {Martinez-Sykora}, {Lites}, {McIntosh}, {Uitenbroek}, {Okamoto}, {Gummin},
  {Auker}, {Jerram}, {Pool}, \& {Waltham}}]{De-Pontieu:2014yu}
{De Pontieu}, B., {Title}, A.~M., {Lemen}, J.~R., {Kushner}, G.~D., {Akin},
  D.~J., {Allard}, B., {Berger}, T., {Boerner}, P., {Cheung}, M., {Chou}, C.,
  {Drake}, J.~F., {Duncan}, D.~W., {Freeland}, S., {Heyman}, G.~F., {Hoffman},
  C., {Hurlburt}, N.~E., {Lindgren}, R.~W., {Mathur}, D., {Rehse}, R.,
  {Sabolish}, D., {Seguin}, R., {Schrijver}, C.~J., {Tarbell}, T.~D.,
  {W{\"u}lser}, J.-P., {Wolfson}, C.~J., {Yanari}, C., {Mudge}, J.,
  {Nguyen-Phuc}, N., {Timmons}, R., {van Bezooijen}, R., {Weingrod}, I.,
  {Brookner}, R., {Butcher}, G., {Dougherty}, B., {Eder}, J., {Knagenhjelm},
  V., {Larsen}, S., {Mansir}, D., {Phan}, L., {Boyle}, P., {Cheimets}, P.~N.,
  {DeLuca}, E.~E., {Golub}, L., {Gates}, R., {Hertz}, E., {McKillop}, S.,
  {Park}, S., {Perry}, T., {Podgorski}, W.~A., {Reeves}, K., {Saar}, S.,
  {Testa}, P., {Tian}, H., {Weber}, M., {Dunn}, C., {Eccles}, S., {Jaeggli},
  S.~A., {Kankelborg}, C.~C., {Mashburn}, K., {Pust}, N., {Springer}, L.,
  {Carvalho}, R., {Kleint}, L., {Marmie}, J., {Mazmanian}, E., {Pereira},
  T.~M.~D., {Sawyer}, S., {Strong}, J., {Worden}, S.~P., {Carlsson}, M.,
  {Hansteen}, V.~H., {Leenaarts}, J., {Wiesmann}, M., {Aloise}, J., {Chu},
  K.-C., {Bush}, R.~I., {Scherrer}, P.~H., {Brekke}, P., {Martinez-Sykora}, J.,
  {Lites}, B.~W., {McIntosh}, S.~W., {Uitenbroek}, H., {Okamoto}, T.~J.,
  {Gummin}, M.~A., {Auker}, G., {Jerram}, P., {Pool}, P., \& {Waltham}, N.
  2014{\natexlab{b}}, \solphys, 289, 2733

\bibitem[{{Golding} {et~al.}(2014){Golding}, {Carlsson}, \&
  {Leenaarts}}]{Golding:2014fk}
{Golding}, T.~P., {Carlsson}, M., \& {Leenaarts}, J. 2014, \apj, 784, 30

\bibitem[{{Gudiksen} {et~al.}(2011){Gudiksen}, {Carlsson}, {Hansteen}, {Hayek},
  {Leenaarts}, \& {Mart{\'{\i}}nez-Sykora}}]{Gudiksen:2011qy}
{Gudiksen}, B.~V., {Carlsson}, M., {Hansteen}, V.~H., {Hayek}, W., {Leenaarts},
  J., \& {Mart{\'{\i}}nez-Sykora}, J. 2011, \aap, 531, A154+

\bibitem[{{Hansteen} {et~al.}(2006){Hansteen}, {De Pontieu}, {Rouppe van der
  Voort}, {van Noort}, \& {Carlsson}}]{Hansteen+DePontieu2006}
{Hansteen}, V.~H., {De Pontieu}, B., {Rouppe van der Voort}, L., {van Noort},
  M., \& {Carlsson}, M. 2006, Apj, 647, L73

\bibitem[{{Hansteen} {et~al.}(2010){Hansteen}, {Hara}, {De Pontieu}, \&
  {Carlsson}}]{Hansteen:2010uq}
{Hansteen}, V.~H., {Hara}, H., {De Pontieu}, B., \& {Carlsson}, M. 2010, \apj,
  718, 1070

\bibitem[{{Hayek} {et~al.}(2010){Hayek}, {Asplund}, {Carlsson}, {Trampedach},
  {Collet}, {Gudiksen}, {Hansteen}, \& {Leenaarts}}]{Hayek:2010ac}
{Hayek}, W., {Asplund}, M., {Carlsson}, M., {Trampedach}, R., {Collet}, R.,
  {Gudiksen}, B.~V., {Hansteen}, V.~H., \& {Leenaarts}, J. 2010, \aap, 517,
  A49+

\bibitem[Judge et al.(2011)]{Judge2011} Judge, P.~G., Tritschler, A., \& Chye Low, B.\ 2011, \apjl, 730, L4 

\bibitem[{{Kubo} {et~al.}(2016){Kubo}, {Katsukawa}, {Suematsu}, {Kano},
  {Bando}, {Narukage}, {Ishikawa}, {Hara}, {Giono}, {Tsuneta}, {Ishikawa},
  {Shimizu}, {Sakao}, {Winebarger}, {Kobayashi}, {Cirtain}, {Champey},
  {Auch{\`e}re}, {Trujillo Bueno}, {Asensio Ramos}, {{\v S}t{\v e}p{\'a}n},
  {Belluzzi}, {Manso Sainz}, {De Pontieu}, {Ichimoto}, {Carlsson}, {Casini}, \&
  {Goto}}]{Kubo:2016lq}
{Kubo}, M., {Katsukawa}, Y., {Suematsu}, Y., {Kano}, R., {Bando}, T.,
  {Narukage}, N., {Ishikawa}, R., {Hara}, H., {Giono}, G., {Tsuneta}, S.,
  {Ishikawa}, S., {Shimizu}, T., {Sakao}, T., {Winebarger}, A., {Kobayashi},
  K., {Cirtain}, J., {Champey}, P., {Auch{\`e}re}, F., {Trujillo Bueno}, J.,
  {Asensio Ramos}, A., {{\v S}t{\v e}p{\'a}n}, J., {Belluzzi}, L., {Manso
  Sainz}, R., {De Pontieu}, B., {Ichimoto}, K., {Carlsson}, M., {Casini}, R.,
  \& {Goto}, M. 2016, \apj, 832, 141

\bibitem[{{Leenaarts} {et~al.}(2007){Leenaarts}, {Carlsson}, {Hansteen}, \&
  {Rutten}}]{Leenaarts:2007sf}
{Leenaarts}, J., {Carlsson}, M., {Hansteen}, V., \& {Rutten}, R.~J. 2007, \aap,
  473, 625

\bibitem[{{Mart{\'{\i}}nez-Sykora} {et~al.}(2012){Mart{\'{\i}}nez-Sykora}, {De
  Pontieu}, \& {Hansteen}}]{Martinez-Sykora:2012uq}
{Mart{\'{\i}}nez-Sykora}, J., {De Pontieu}, B., \& {Hansteen}, V. 2012, \apj,
  753, 161

\bibitem[{{Mart{\'{\i}}nez-Sykora}
  {et~al.}(2017{\natexlab{a}}){Mart{\'{\i}}nez-Sykora}, {De Pontieu},
  {Hansteen}, \& {Carlsson}}]{Martinez2017b}
{Mart{\'{\i}}nez-Sykora}, J., {De Pontieu}, B., {Hansteen}, V.~H., \&
  {Carlsson}, M. 2017{\natexlab{a}}, ApJ, submitted to

\bibitem[{{Mart{\'{\i}}nez-Sykora}
  {et~al.}(2017{\natexlab{b}}){Mart{\'{\i}}nez-Sykora}, {De Pontieu},
  {Hansteen}, {Rouppe van der Voort}, {Carlsson}, \& {Pereira}}]{Martinez2017}
{Mart{\'{\i}}nez-Sykora}, J., {De Pontieu}, B., {Hansteen}, V.~H., {Rouppe van
  der Voort}, L.~H.~M., {Carlsson}, M., \& {Pereira}, T.~M.~D.
  2017{\natexlab{b}}, Science, 356, 1269

\bibitem[{{Mart{\'{\i}}nez-Sykora} {et~al.}(2009){Mart{\'{\i}}nez-Sykora},
  {Hansteen}, {De Pontieu}, \& {Carlsson}}]{Martinez2009}
{Mart{\'{\i}}nez-Sykora}, J., {Hansteen}, V., {De Pontieu}, B., \& {Carlsson},
  M. 2009, \apj, 701, 1569

\bibitem[{{McIntosh} {et~al.}(2011){McIntosh}, {de Pontieu}, {Carlsson},
  {Hansteen}, {Boerner}, \& {Goossens}}]{McIntosh:2011fk}
{McIntosh}, S.~W., {de Pontieu}, B., {Carlsson}, M., {Hansteen}, V., {Boerner},
  P., \& {Goossens}, M. 2011, \nat, 475, 477

\bibitem[{{Narang} {et~al.}(2016){Narang}, {Arbacher}, {Tian}, {Banerjee},
  {Cranmer}, {DeLuca}, \& {McKillop}}]{Narang2016}
{Narang}, N., {Arbacher}, R.~T., {Tian}, H., {Banerjee}, D., {Cranmer}, S.~R.,
  {DeLuca}, E.~E., \& {McKillop}, S. 2016, \solphys, 291, 1129

\bibitem[{{Pereira} {et~al.}(2012){Pereira}, {De Pontieu}, \&
  {Carlsson}}]{Pereira2012}
{Pereira}, T.~M.~D., {De Pontieu}, B., \& {Carlsson}, M. 2012, \apj, 759, 18

\bibitem[{{Raouafi} {et~al.}(2016){Raouafi}, {Patsourakos}, {Pariat}, {Young},
  {Sterling}, {Savcheva}, {Shimojo}, {Moreno-Insertis}, {DeVore}, {Archontis},
  {T{\"o}r{\"o}k}, {Mason}, {Curdt}, {Meyer}, {Dalmasse}, \&
  {Matsui}}]{Raouafi:2016fe}
{Raouafi}, N.~E., {Patsourakos}, S., {Pariat}, E., {Young}, P.~R., {Sterling},
  A.~C., {Savcheva}, A., {Shimojo}, M., {Moreno-Insertis}, F., {DeVore}, C.~R.,
  {Archontis}, V., {T{\"o}r{\"o}k}, T., {Mason}, H., {Curdt}, W., {Meyer}, K.,
  {Dalmasse}, K., \& {Matsui}, Y. 2016, \ssr, 201, 1

\bibitem[{{Rouppe van der Voort} {et~al.}(2015){Rouppe van der
  Voort}, {De Pontieu}, {Pereira}, {Carlsson}, \& {Hansteen}}]{Rouppe2015}
{Rouppe van der Voort}, L., {De Pontieu}, B., {Pereira}, T.~M.~D., {Carlsson},
  M., \& {Hansteen}, V. 2015, \apjl, 799, L3

\bibitem[{{Rouppe van der Voort} {et~al.}(2007){Rouppe van der Voort}, {De
  Pontieu}, {Hansteen}, {Carlsson}, \& {van Noort}}]{luc2007}
{Rouppe van der Voort}, L.~H.~M., {De Pontieu}, B., {Hansteen}, V.~H.,
  {Carlsson}, M., \& {van Noort}, M. 2007, \apjl, 660, L169

\bibitem[{{Sekse} {et~al.}(2012){Sekse}, {Rouppe van der Voort}, \& {De
  Pontieu}}]{Sekse2012}
{Sekse}, D.~H., {Rouppe van der Voort}, L., \& {De Pontieu}, B. 2012, \apj,
  752, 108

\bibitem[{{Sekse} {et~al.}(2013){Sekse}, {Rouppe van der Voort}, \& {De
  Pontieu}}]{Sekse2013}
---. 2013, \apj, 764, 164

\bibitem[Skogsrud et al.(2015)]{Skogsrud2015} Skogsrud, H., 
Rouppe van der Voort, L., De Pontieu, B., \& Pereira, T.~M.~D.\ 2015, \apj, 806, 170 

\bibitem[Skogsrud et al.(2016)]{Skogsrud2016} Skogsrud, H., Rouppe van der Voort, L., \& De Pontieu, B.\ 2016, \apj, 817, 124 

\bibitem[{{Sterling}(2000)}]{Sterling2000}
{Sterling}, A.~C. 2000, \solphys, 196, 79

\bibitem[{{Tian} {et~al.}(2014){Tian}, {DeLuca}, {Cranmer}, {De Pontieu},
  {Peter}, {Mart{\'{\i}}nez-Sykora}, {Golub}, {McKillop}, {Reeves}, {Miralles},
  {McCauley}, {Saar}, {Testa}, {Weber}, {Murphy}, {Lemen}, {Title}, {Boerner},
  {Hurlburt}, {Tarbell}, {Wuelser}, {Kleint}, {Kankelborg}, {Jaeggli},
  {Carlsson}, {Hansteen}, \& {McIntosh}}]{Tian:2014fp}
{Tian}, H., {DeLuca}, E.~E., {Cranmer}, S.~R., {De Pontieu}, B., {Peter}, H.,
  {Mart{\'{\i}}nez-Sykora}, J., {Golub}, L., {McKillop}, S., {Reeves}, K.~K.,
  {Miralles}, M.~P., {McCauley}, P., {Saar}, S., {Testa}, P., {Weber}, M.,
  {Murphy}, N., {Lemen}, J., {Title}, A., {Boerner}, P., {Hurlburt}, N.,
  {Tarbell}, T.~D., {Wuelser}, J.~P., {Kleint}, L., {Kankelborg}, C.,
  {Jaeggli}, S., {Carlsson}, M., {Hansteen}, V., \& {McIntosh}, S.~W. 2014,
  Science, 346, A315

\bibitem[{{Tsiropoula} {et~al.}(2012){Tsiropoula}, {Tziotziou}, {Kontogiannis},
  {Madjarska}, {Doyle}, \& {Suematsu}}]{Tsiropoula2012}
{Tsiropoula}, G., {Tziotziou}, K., {Kontogiannis}, I., {Madjarska}, M.~S.,
  {Doyle}, J.~G., \& {Suematsu}, Y. 2012, \ssr, 169, 181

\end{thebibliography}

\end{document}